\documentclass[12pt]{iopart}

\usepackage{graphicx}
  
\begin{document}

\title[Pair distribution functions of the 2DEG
with two symmetric valleys]{Pair distribution functions of the two-dimensional
electron gas with two symmetric valleys}

\author{M Marchi$^1$$^,$$^2$, S De Palo$^1$$^,$$^3$, S Moroni$^1$$^,$$^2$ 
and  G Senatore$^1$$^,$$^3$}
\address{$^1$ INFM-CNR DEMOCRITOS National
Simulation Center, Trieste, Italy}
\address{$^2$ SISSA, International School
for Advanced Studies, via Beirut 2-4, 34014 Trieste, Italy}
\address{$^3$ Dipartimento di Fisica
Teorica, Universit\`a di Trieste, Strada Costiera 11, 34014
Trieste, Italy}
\ead{marchi@sissa.it}

\begin{abstract}
We present component-resolved and total 
pair distribution functions for a 2DEG with two symmetric valleys. Our results
are based on quantum Monte Carlo simulations performed at several
densities and spin polarizations. 

\end{abstract}


\section{Introduction}
A  two-valley (2V) two-dimensional (2D) electron gas (EG) is the
simplest model to describe electrons confined in solid state devices
such as Si-MOSFETs  \cite{ando} and certain quantum wells \cite{shkolnikov}. 
In this 2D model,
electrons interact via a $1/r$ potential in a uniform
neutralizing background and possess an additional discrete degree of freedom 
(the valley index).
One may identify electrons with given spin projection and valley index as 
belonging to a different species or \emph{component}. 
We focus on the case of two symmetric valleys, where the number
of up (down) spin electrons is the same for both valleys. In this case 
at zero temperature
the 2DEG is completely
characterized in terms of the  coupling parameter $r_s=1/\sqrt{\pi\,n}\,a_B$
and the spin polarization $\zeta=(n_{\uparrow}-n_{\downarrow})/n$ (where $n$ is 
the total electron density, $a_B$ the Bohr radius, 
$n_{\uparrow(\downarrow)}$ the density of up (down) spin electrons). 

The interest in the properties of the 2DEG has been
strongly revived in the last years due to the experimental
discovery of a previously unexpected metal-insulator transition 
 \cite{krav-rev,punnoose_nature} in which the valley degree of freedom
appears to play an important role  \cite{punnoose,spinvalley}. The
transition takes place at low density, where an accurate treatment of
electron correlation is crucial. In this
respect, quantum Monte Carlo (QMC) simulations have provided over the
years the method of choice for microscopic studies
 \cite{cep89,kwon93,rapisarda,conti,varsano,attacca,pgg_pd} of the 2DEG. 

We have recently provided an analytic expression of the correlation energy 
of the 2V2DEG \cite{marchi_2v}.
Here, we focus on the pair distribution functions (PDFs), which
are strictly related to the description of exchange and correlation properties 
of the system (see \emph{e.g.} Ref.~\cite{GV}).
As pointed out in Ref.~\cite{pgg_pd}, PDFs may serve a variety
of purposes, among which the estimate of finite thickness effects on the 2DEG 
spin susceptibility \cite{depalo05}, applications in DFT calculations, or a test
of the accuracy of hypernetted-chain calculations \cite{dharmawardana}.
\section{Pair distribution function: definitions}
\label{pdf_def}
The PDF  is related to the probability of
finding two electrons at positions $\mathbf{r}$ and $\mathbf{r'}$ respectively.
The \emph{component-resolved} PDF
$g_{\alpha\beta}(\mathbf{r'},\mathbf{r})$ of a multicomponent 2DEG is
defined as \cite{GV}
\begin{eqnarray}
\label{sv_pdf}
g_{\alpha\beta}(\mathbf{r'},\mathbf{r})=
\frac{\langle \psi^{\dagger}_{\beta}(\mathbf{r'})
\psi^{\dagger}_{\alpha}(\mathbf{r})\psi_{\alpha}(\mathbf{r})
\psi_{\beta}(\mathbf{r'})
\rangle}{n_{\beta}(\mathbf{r'})n_{\alpha}(\mathbf{r})},
\end{eqnarray}
with $\psi^{\dagger}_{\alpha}$, $\psi_{\alpha}$ denoting creation and 
annihilation field operators, $\langle\dots\rangle$ the expectation value on 
the ground state and $n_{\alpha}(\mathbf{r})=
\langle \psi^{\dagger}_{\alpha}(\mathbf{r})\psi_{\alpha}(\mathbf{r})\rangle$ 
the electron density of the component $\alpha$.
The normalization is such
that $g_{\alpha\beta}\equiv 1$, in case  there is neither exchange nor
correlation. 
In a homogeneous and isotropic system, $g_{\alpha\beta}$ depends only on the
relative distance $r=|\mathbf{r}-\mathbf{r'}|$ and there is symmetry for
index permutations ($g_{\alpha\beta}=g_{\beta\alpha}$).
If $c_{\alpha}=n_{\alpha}/n$ is the concentration of the component 
$\alpha$, the total (\emph{component-summed}) PDF $g(r)$ reads
\begin{eqnarray}
\label{tot_pdf}
g(r)=\sum_{\alpha\beta}c_{\alpha}c_{\beta}\,g_{\alpha\beta}(r).
\end{eqnarray}

In the 2V2DEG, 
$\alpha\equiv\sigma\nu$ is a composite index  which 
denotes the spin and valley (respectively $\sigma$ and $\nu$) degrees of 
freedom and spans the four cases
$\alpha=\uparrow 1,\downarrow 1, \uparrow 2, \downarrow 2$.
In general, for 2V there are ten different $g_{\alpha\beta}$, but in the case 
of two symmetric valleys the number of different $g_{\alpha\beta}$ is two
for $\zeta=0,1$ and five for $0<\zeta< 1$ (see also Ref.~\cite{dharmawardana}).
In the following we shall denote the different $g_{\alpha\beta}$ with one
of the possible different labels (\emph{e.g.} for $0<\zeta< 1$ 
$g_{\uparrow 1\downarrow 1}= g_{\uparrow 2\downarrow 2}=
  g_{\uparrow 1\downarrow 2}= g_{\uparrow 2\downarrow 1}$).
\section{Simulation details}
\label{dmc}
Most of the simulation details are the same as in Ref.~\cite{marchi_2v}. We
performed fixed-phase DMC simulations (for a review of QMC techniques 
see \emph{e.g.} Ref~\cite{foulkes}) for $r_s=1,2,5,10,20$ and 
$\zeta=0,3/13,5/13,7/13,10/13,1$. 
To reduce the finite size effects,
we used twist-averaged boundary conditions (TABC) 
\cite{tabc}, which also allow to change $\zeta$ by flipping
any number of spins at fixed number of electrons $N$. To sample the 
PDFs we performed simulations for a system of  $N=52$ electrons.
Time steps were chosen at the different $r_s$ to give an acceptance rate 
corresponding to $\sim 99\%$. We did simulations with 320 walkers. 
The twist grid is the same as in Ref.~\cite{marchi_2v}.
As in Ref.~\cite{marchi_2v}, we used a Slater-Jastrow trial wave function 
$\Psi_T$, but, here, we considered only plane-wave nodes, 
since the more accurate backflow (BF) nodes 
yield only slight modifications to the 
PDFs \cite{kwon93,pgg_pd}. Besides, BF effects on the energy were
shown to be bigger in the two-component case than in the four-component system
\cite{marchi_2v}.

DMC provides the mixed estimate of an operator $O$, \emph{i.e.}
$\langle O \rangle_{mix}=\langle \Psi_0|O|\Psi_T\rangle/
\langle\Psi_0\Psi_T\rangle$ (with $\Psi_0$ denoting the ground state of the
system). If $O$ commutes with the Hamiltonian $H$, 
$\langle O \rangle_{mix}$ coincides with the expectation
value on the true ground state $\Psi_0$.
If $O$ does not commute with $H$ (as in the case here considered), it is 
better to compute the extrapolated estimate 
$\langle O \rangle_{extr} =2\langle O\rangle_{mix}-\langle O\rangle_{VMC} + 
{{O(\delta^2)}}$ (where $\langle O\rangle_{VMC}$ is the variational MC 
expectation value on $\Psi_T$)
which is accurate to second order in the difference $\delta$
between $\Psi_0$ and $\Psi_T$.
\section{Two valley pair distribution functions}
\label{figure_results}
\begin{figure}[h]
\begin{center}
\includegraphics{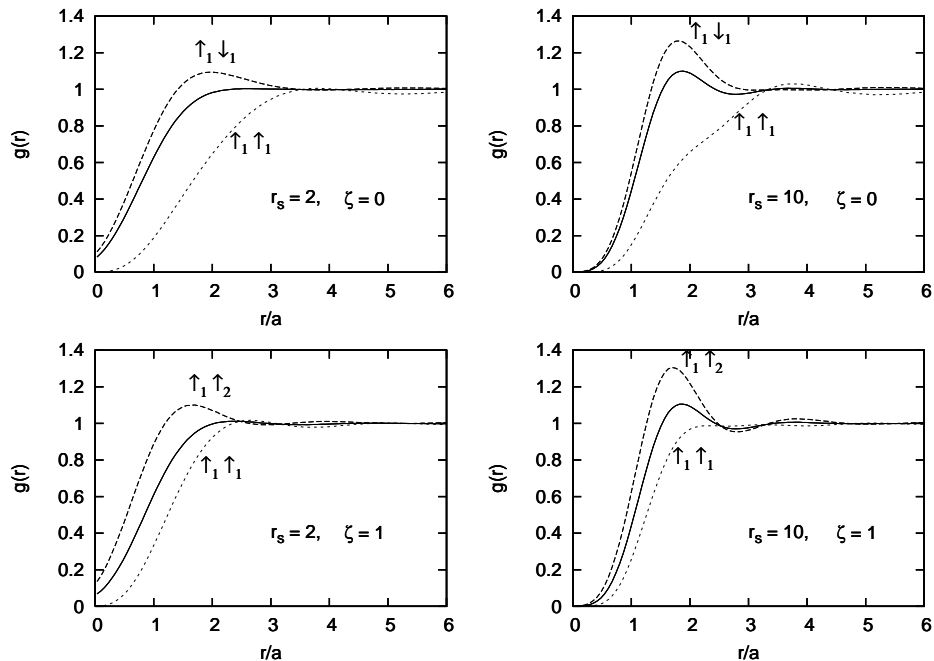}
\caption{2V component-resolved pair distribution functions for 
$\zeta=0,\,1$ and $r_s=2,10$. Solid lines represent the total $g(r)$.}
\label{2vspinresolved-z01}
\end{center}
\end{figure}
\begin{figure}[h]
\begin{center}
\includegraphics{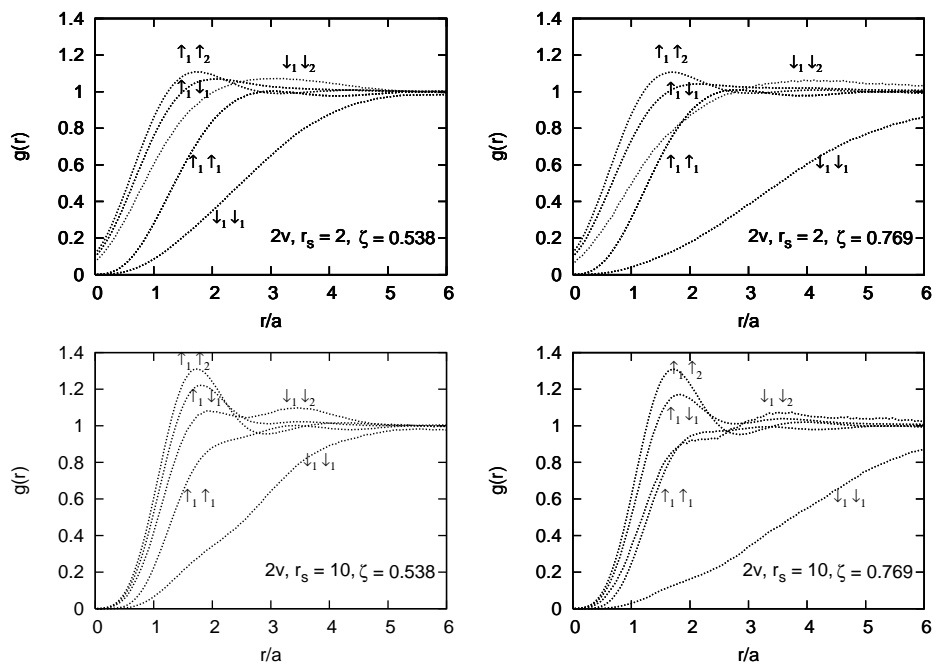}
\caption{Examples of 2V component resolved pair distribution functions 
at finite  $\zeta$. See labels.}
\label{2vspinresolved-zno01}
\end{center}
\end{figure}
\begin{figure}[h]
\begin{center}
\includegraphics{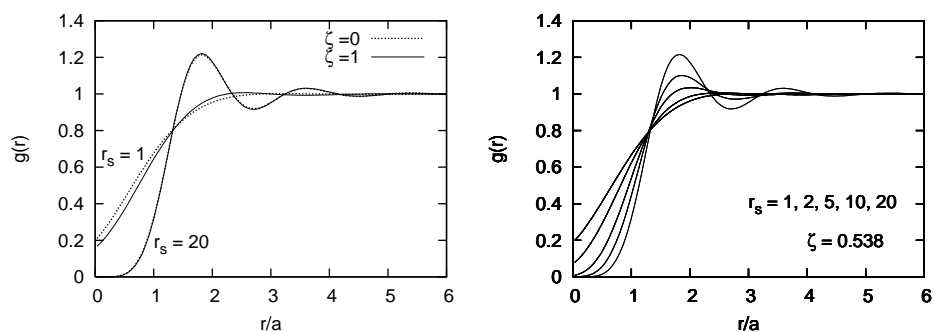}
\caption{2V total pair distribution functions. Left panel: $r_s=1,20$ and
$\zeta=0,1$. Right panel: 
$r_s=1,\,2,\,5,\,10,\,20$ and $\zeta=0.538$ (increasing peaks for
increasing $r_s$). All lengths are in units of $a=r_s a_B$.}
\label{2vspinsummed}
\end{center}
\end{figure}

 
In Fig.~\ref{2vspinresolved-z01}-\ref{2vspinresolved-zno01}
we show representative examples of 2V component-resolved PDFs
for $\zeta=0,1$ and finite $\zeta$ respectively.
All lenghts are given in units of $a=r_s a_B$.
The component-resolved PDFs shown in Fig.~\ref{2vspinresolved-z01} illustrate 
the tendency to a local order which favors electrons belonging to different 
species (as e.g. $\uparrow 1$ and $\downarrow 1$ ) to get closer than electrons 
belonging to the same species.
For intermediate spin polarizations (see Fig.~\ref{2vspinresolved-zno01})
the component-resolved PDFs exhibit a
richer structure than the $\zeta=0,1$ cases, with qualitative features
clearly related to the interplay of exchange and correlations (for example,
in the density range considered the
diagonal PDF of the minority component in a strongly polarized system
is found to be determined by exchange alone, to a very good approximation).
The significant spin-polarization dependence seen in the component-resolved
PDFs almost disappears in the total PDFs, particularly at large $r_s$. This can
be appreciated in Fig.~\ref{2vspinsummed} (left panel), which shows the total 
PDFs for a high density ($r_s=1$) and a low density ($r_s=20$) case 
with zero and full spin polarization.
The dependence of the total PDF on $r_s$ is depicted in Fig.~\ref{2vspinsummed} 
for  $\zeta=0.538$.

We see that even for $r_s$ as small as 1 the effect of exchange on the
total PDF, which is
expected to decrease with the number $N_c$ of (equally populated) components,
is already very small for $N_c=2$.

Full tabulations of the calculated PDFs are available upon request
from the first author.

\end{document}